\documentclass[10pt,conference]{IEEEtran}

\usepackage[utf8]{inputenc}
\usepackage[T1]{fontenc}

\usepackage{xcolor}
\usepackage{amsmath}
\usepackage{amssymb}
\usepackage{graphicx}
\usepackage{enumitem}  
\usepackage[export]{adjustbox}
\usepackage{tikz}
\usetikzlibrary{quantikz}

\begin{document}

\title{Teaching Quantum Computing using Microsoft Quantum Development Kit and Azure Quantum}
\author{
  \IEEEauthorblockN{Mariia Mykhailova}
  \IEEEauthorblockA{mamykhai@microsoft.com \\ Microsoft Quantum, United States}
}

\IEEEoverridecommandlockouts \IEEEpubid{\begin{minipage}{\textwidth}\ \\[12pt]\ \copyright2023 IEEE. Personal use of this material is permitted. Permission from IEEE must be obtained for all other uses, in any current or future media, including reprinting/republishing this material for advertising or promotional purposes, creating new collective works, for resale or redistribution to servers or lists, or reuse of any copyrighted component of this work in other works.\hfill \end{minipage}}

\maketitle

\begin{abstract}
This report describes my experience teaching a graduate-level quantum computing course at Northeastern University in the academic year 2022--23.
The course takes a practical, software-driven approach to the course, teaching basic quantum concepts and algorithms through hands-on programming assignments and a software-focused final project.
The course guides learners through all stages of the quantum software development process, from solving quantum computing problems and implementing solutions to debugging quantum programs, optimizing the code, and running the code on quantum hardware.
This report offers instructors who want to adopt a similar practical approach to teaching quantum computing a comprehensive guide to getting started.
\end{abstract}

\section{Introduction}

Quantum computing represents a novel approach to computing, utilizing quantum-mechanical phenomena like superposition and entanglement to execute certain computational tasks more efficiently compared to classical computing.
The increasing interest and financial support in the field of quantum information science and engineering (QISE) have resulted in a growing need for quantum-trained workforce.

Recent evaluations of the needs of quantum industry\cite{QEDC-workforce-assessment}\cite{Aiello_2021} identified quantum software engineering and application development as one of the competencies essential for certain roles within the sector.
Surveys of the university programs offering Master-level education in QISE\cite{Aiello_2021}\cite{Frantz_2020} show that quantum programming is often incorporated in them as an independent course or as a component within an introductory course.

In this report I describe my experience teaching a graduate course ``CSYE6305: Introduction to Quantum Computing with Applications'' at Northeastern University during the fall and spring semesters of the academic year 2022--23.
I outline the guiding principles behind the course design, describe the programming assignments created for the course, and discuss the lessons learned.

I hope that this work will inspire more instructors to adopt a similar software-driven approach to delivering quantum computing courses and enable a broader student audience to learn quantum programming.

%=============================================================================
\section{Curriculum and course structure}

The course targets engineering graduate students who do not specialize in quantum information science and do not necessarily have an extensive background in physics, mathematics, or theoretical computer science. 
The students' background means that traditional approaches to teaching quantum computing, focused either on the mathematical presentation of the field or on the theoretical physics of quantum systems and hardware devices, would be inefficient.

\IEEEpubidadjcol

The course approached the subject of quantum computing from a computer science point of view, similar to \cite{UW}.
It focused on teaching basic concepts and algorithms of quantum computing in a practical manner, requiring students to apply their theoretical knowledge to solving problems, write the quantum code to implement the solutions, and verify its correctness using quantum simulators or explore its behavior when executed on quantum hardware.

The course covered a combination of introductory topics featured in most courses and textbooks on quantum computing with deeper dives into the topics important for understanding the current landscape and the future directions of quantum computing.
It consisted of $10$ week-long lecture modules (the number selected to match the length of the semester), an introductory module covering the required software setup and the math prerequisites (linear algebra tutorial), and a final project. The lecture modules were, in order of presentation:

\begin{enumerate}
    \item Single-qubit quantum systems: the concept of a quantum state, superposition, single-qubit gates, measurements.
    \item Multi-qubit quantum systems: multi-qubit quantum states, entanglement, multi-qubit gates, measurements of multi-qubit systems (including partial measurements).
    \item Simple communication algorithms: BB84 quantum key distribution algorithm, teleportation, superdense coding.
    \item Quantum phase oracles and simple oracular algorithms: Deutsch, Deutsch-Josza, and Bernstein-Vazirani.
    \item Reversible computing: reversible Boolean logic, reversible circuit synthesis, implementing marking oracles.
    \item Grover's search algorithm and using it to solve problems.
    \item Quantum software stack.
    \item Building up to Shor’s algorithm: Fourier transform, phase estimation, Shor’s algorithm for integer factorization.
    \item Quantum error correction and fault-tolerant quantum computing.
    \item The current landscape of quantum hardware development and the quantum community.
\end{enumerate}

Each of the lecture modules included a 3-hour lecture that introduced the relevant concepts, explained quantum algorithms, and covered the software tools necessary for completing the module's assignments. 
The supplementary materials for the modules included recommended and optional reading material, and tutorials and katas from the open-source project Quantum Katas \cite{quantum-katas} for hands-on problem solving and programming practice. 

Student performance was evaluated in two ways:

\begin{itemize}
    \item weekly programming assignments (60\% of the final grade), and
    \item the final project (40\% of the final grade in fall 2022, up to 65\% of the final grade in spring 2023).
\end{itemize}

Weekly programming assignments were offered in the introductory module and the first six lecture modules, up to and including module 6 (``Grover's search algorithm''). Most modules included multiple types of assignments; for example, module 1 (``Single-qubit quantum systems'') included automatically graded programming assignments on single-qubit quantum systems and a hardware exploration assignment in which students explored running a quantum random bit generator on a cloud simulator and on a quantum device.

The final project was introduced after module 7 (``Quantum software stack''). Students worked on it during the last month of the course and presented their work at the last meeting.

The course was taught using Q\#\cite{rwdsl2018}, a high-level domain-specific quantum programming language, and the Microsoft Quantum Development Kit (QDK)\footnote{https://learn.microsoft.com/azure/quantum/overview-what-is-qsharp-and-qdk}, an open-source software development kit that includes a Q\# compiler, a variety of quantum simulators, and other tools for quantum software design and development.
The assignments that required running quantum programs on quantum hardware used Azure Quantum\footnote{https://learn.microsoft.com/azure/quantum/overview-azure-quantum}, a cloud service that provides access to quantum hardware and simulators from different companies. The use of Azure Quantum in the course was covered by the Azure Quantum Credits program\footnote{https://learn.microsoft.com/azure/quantum/azure-quantum-credits} and was free for the students, which allowed them to experiment with running their programs on the hardware without worrying about the costs.

%=============================================================================
\section{Automatically graded quantum programming problems}

Most of the assignments in the course required students to apply the theory they learned in the module to solve small, practical programming problems related to the topic.
The tasks in these assignments followed the structure of the problems in the Quantum Katas\cite{quantum-katas}. 
Each task described a specific quantum computing problem and provided the signature of a Q\# operation that needed to be implemented to solve the given problem. 
Students then had to fill the body of the operation with the code that implemented the solution.

Similar to the problems in the Quantum Katas, programming assignments of this type supported automatic grading and student self-evaluation. 
Automatic grading was implemented using predefined testing harnesses for the tasks - Q\# projects that ran the solutions on a set of tests and validated that the results produced by the students' code matched the expected ones. 
The testing harnesses relied on the use of quantum simulators and program validation tools available in the QDK, described in more detail in \cite{MS21}.
Examples of the tasks that were offered in automatically graded assignments included:
\begin{itemize}
    \item preparing the described quantum state,
    \item performing the measurements to identify the quantum state of the given qubits,
    \item implementing the required unitary transformation, for example, a quantum oracle implementing the given classical function, and others.
\end{itemize}

This approach enabled the automation of the grading process, significantly reducing the workload on the instructor.
Additionally, sharing the testing harnesses with the students as part of the assignment allowed them to receive feedback on their solutions before submitting them for grading.

\begin{figure}
    \centering
    \includegraphics[width=.95\linewidth]{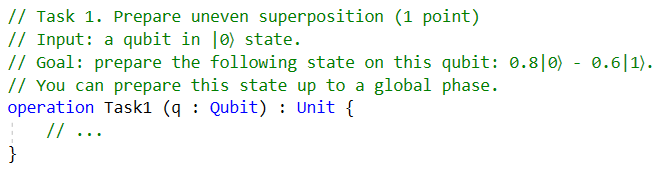}
    \caption{State preparation task that asks the student to replace the ``\texttt{// ...}'' comment with Q\# code that prepares the qubit in the described state}
    \label{fig:state-prep-task}
\end{figure}

Figure~\ref{fig:state-prep-task} shows an example of a task that asks the student to prepare a quantum state $0.8|0\rangle - 0.6|1\rangle$. 
The corresponding Q\# operation takes one input parameter - the qubit in the $|0\rangle$ state - and does not produce an output. 
Since it should prepare a quantum state, it acts by changing the state of the qubit passed to it as the argument rather than by producing an output.

The testing harness for this task used a state vector simulator (the full state simulator or the sparse simulator included in the QDK\footnote{https://learn.microsoft.com/azure/quantum/machines/}). 
It applied the student's solution to a qubit in the $|0\rangle$ state, followed by applying the adjoint of the ``reference'' solution - the instructor's solution known to be correct. 
If the student's solution prepared the expected state, this sequence of steps resulted in the qubit returning to the $|0\rangle$ state; otherwise, the qubit would end up in a different state. 
The state of the qubit at the end of the test execution was validated using built-in QDK tools \cite{MS21}.

%=============================================================================
\section{Debugging quantum programs}

Assignments that require the students to identify and fix the issues in the given quantum programs aim to enhance both their proficiency with the programming tools employed in the course and their understanding of the algorithms studied in it.

We can distinguish three types of errors found in programs (both classical and quantum) based on the ways they manifest.

\begin{description}
    \item[Syntax errors] Errors in the syntax of the program that prevent the code from being recognized by the compiler or the interpreter. Since Q\# is a compiled language, syntax errors are typically detected at compile-time or even earlier, when the code is first opened in an IDE such as Visual Studio Code.
    \item[Runtime errors] Errors in the program logic that manifest as exceptions thrown during program execution. Examples of runtime errors include classical errors, such as attempting to access an array element using an index outside of array bounds, and quantum-specific errors, such as using the same qubit as both the control and the target for a controlled gate. Identifying runtime errors requires fixing all syntax errors, running the program to obtain the exception, and tracing it back to its source.
    \item[Logical errors] Errors in the program logic that do not prevent the program from compiling and executing but produce incorrect results. Detecting logical errors requires fixing all syntax and runtime errors, executing the program to get its results, analyzing these results for accuracy, and identifying the underlying reasons for incorrect results. In the case of probabilistic algorithms or errors that appear intermittently (such as using the wrong basis to perform the final measurements, which may occasionally produce correct results), identifying the errors might require more advanced statistical analysis. However, often it suffices to run the algorithm several times to notice the presence of an error.
\end{description}

I used this kind of assignments in the modules that covered the BB84 quantum key distribution protocol and Grover's search algorithm. 
They allow offering both easier tasks that only check the students' familiarity with the programming language (syntax and runtime errors are typically easy to find and fix) and more advanced tasks that verify their understanding of the end-to-end logic of the algorithm and its expected behavior.

\begin{figure}
    \centering
    \includegraphics[width=.98\linewidth]{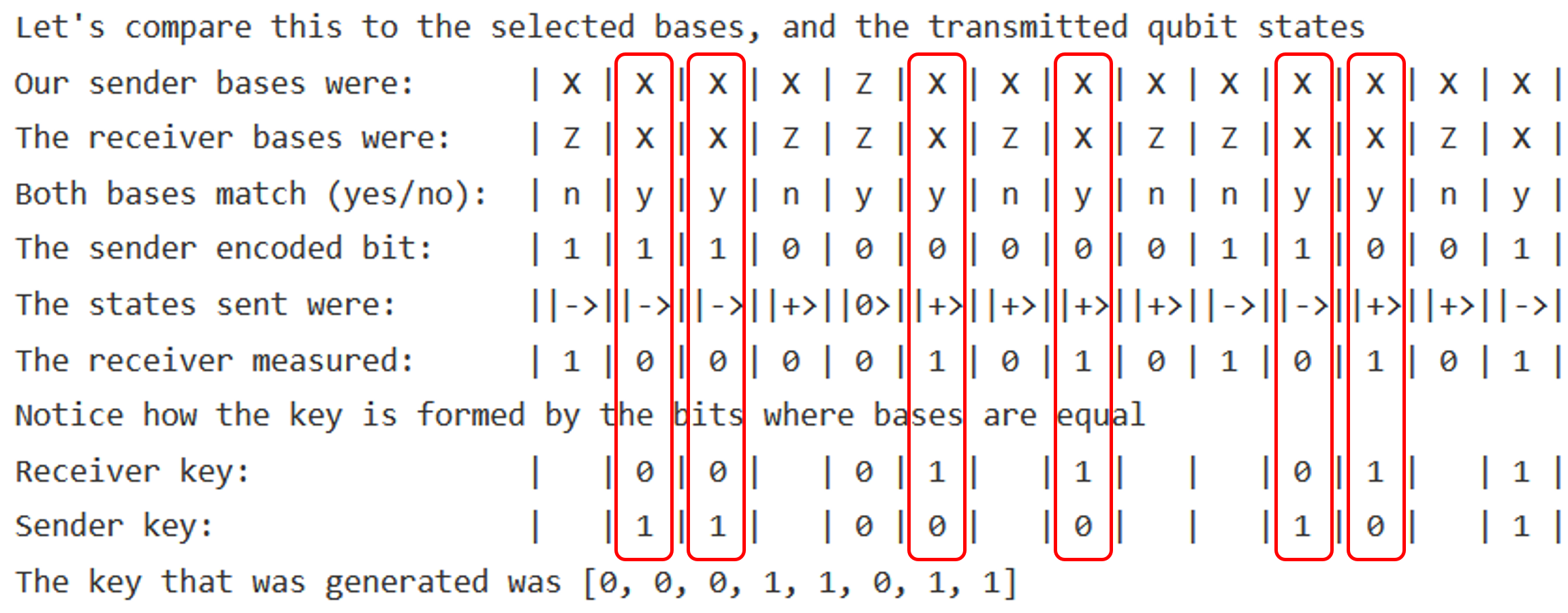}
    \caption{Debugging task that shows the output of the BB84 protocol with a logical error (incorrect gate used to change the basis of the qubit on the sender side)}
    \label{fig:debugging-task}
\end{figure}

Figure~\ref{fig:debugging-task} shows an example output of a program offered in a debugging assignment, the BB84 protocol executed in a noiseless environment without an eavesdropper. The highlighted columns show the cases in which the receiver's measurement results do not match the sender's bit when both of them selected the X basis, and the last column shows that sometimes the bits still match. This discrepancy indicates a logical error when processing the X basis either in the sender's encoding step or in the receiver's decoding step, but not simply an error in labeling the bits (which would show up every time both parties selected the X basis), which likely means an error in the basis used for encoding or measurement.

%=============================================================================
\section{Optimizing quantum programs}

The next type of assignments introduced students to two other important steps of the quantum software development cycle - resource estimation and program optimization. 

Resource estimation tools provide accurate automated estimates of the resources (typically the total number of qubits and the circuit depth, sometimes represented as the time required to execute the program) required to run the given quantum program on a quantum device, even if the program is too large to be executed or simulated. 
This assignment used Azure Quantum Resource Estimator, a cloud service that estimated the resources required to implement a given quantum algorithm on a specified architecture of a digital quantum computer with the specified underlying qubit technology\cite{beverland2022assessing}.

In this course, resource estimation was introduced in the module on reversible computing as part of a circuit optimization task. Resource estimation tools can also be useful for more advanced topics, such as discussions of different approaches to solving the same problem or different implementations of the same arithmetic routine, since they allow students to easily compare the resources required by different programs. Furthermore, the same tools can be used to illustrate the overhead introduced by error correction and the impact of the choice of error correction scheme on the resource requirements of the algorithm. 

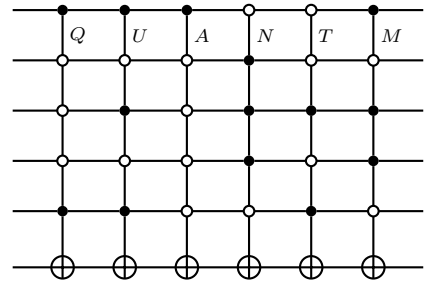
\begin{figure}
    \centering
    \begin{quantikz}
    & \ctrl{1}{Q}  & \ctrl{1}{U}  & \ctrl{1}{A}  & \octrl{1}{N} & \octrl{1}{T} & \ctrl{1}{M}  & \qw \\
    & \octrl{1} & \octrl{1} & \octrl{1} & \ctrl{1}  & \octrl{1} & \octrl{1} & \qw \\
    & \octrl{1} & \ctrl{1}  & \octrl{1} & \ctrl{1}  & \ctrl{1}  & \ctrl{1}  & \qw \\
    & \octrl{1} & \octrl{1} & \octrl{1} & \ctrl{1}  & \octrl{1} & \ctrl{1}  & \qw \\
    & \ctrl{1}  & \ctrl{1}  & \octrl{1} & \octrl{1} & \ctrl{1}  & \octrl{1} & \qw \\
    & \targ{}   & \targ{}   & \targ{}   & \targ{}   & \targ{}   & \targ{}   & \qw
    \end{quantikz}
    \caption{Circuit optimization task that asks the student to rewrite the circuit to minimize the resources required to run it. The control bit sequences of the gates spell the alphabet positions of letters QUANTM (the top wire is the least significant bit of the letter number).}
    \label{fig:circuit-optimization-task}
\end{figure}

Figure~\ref{fig:circuit-optimization-task} shows an example of a task that asks the student to rewrite the given code to minimize the resources required to run it while preserving the computation it performs. 
The metric to be minimized was defined as (the number of logical algorithmic qubits) * (the algorithmic depth of the circuit). This metric uses the logical depth and width of the circuit, taking into account the additional qubits and circuit depth required to implement multi-controlled gates, but not drilling down into the physical resource requirements introduced by error correction and T-factories, which is an appropriate level of detail at that point in the course.

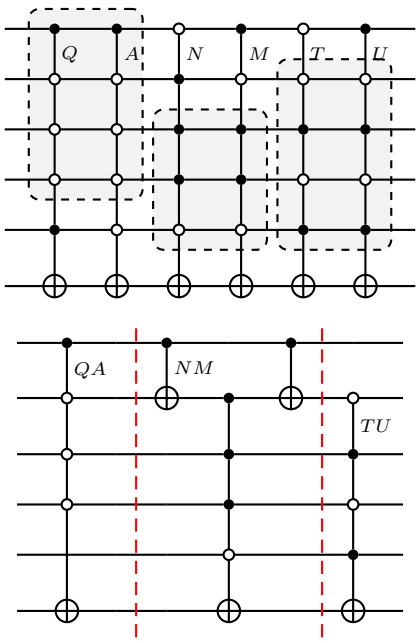
\begin{figure}
    \centering
    \begin{quantikz}
    & \ctrl{1}{Q}\gategroup[wires = 4,steps=2,style={dashed,
rounded corners,inner sep=2pt,fill=gray!10},background]{}  & \ctrl{1}{A} & \octrl{1}{N} & \ctrl{1}{M} & \octrl{1}{T} & \ctrl{1}{U}  & \qw \\
    & \octrl{1} & \octrl{1} & \ctrl{1}  & \octrl{1} & \octrl{1}\gategroup[wires = 4,steps=2,style={dashed,
rounded corners,inner sep=2pt,fill=gray!10},background]{} & \octrl{1} & \qw \\
    & \octrl{1} & \octrl{1} & \ctrl{1}\gategroup[wires = 3,steps=2,style={dashed,
rounded corners,inner sep=2pt,fill=gray!10},background]{}  & \ctrl{1}  & \ctrl{1}  & \ctrl{1}  & \qw \\
    & \octrl{1} & \octrl{1} & \ctrl{1}  & \ctrl{1}  & \octrl{1} & \octrl{1} & \qw \\
    & \ctrl{1}  & \octrl{1} & \octrl{1} & \octrl{1} & \ctrl{1}  & \ctrl{1}  & \qw \\
    & \targ{}   & \targ{}   & \targ{}   & \targ{}   & \targ{}   & \targ{}   & \qw
    \end{quantikz}
    
    \begin{quantikz}
    & \ctrl{1}{QA} & \qw\slice{} & \ctrl{1}{NM} & \qw       & \ctrl{1}\slice{} & \qw       & \qw \\
    & \octrl{1}    & \qw         & \targ{}  & \ctrl{1}  & \targ{}          & \octrl{1}{TU} & \qw \\
    & \octrl{1}    & \qw         & \qw      & \ctrl{1}  & \qw              & \ctrl{1}  & \qw \\
    & \octrl{2}    & \qw         & \qw      & \ctrl{1}  & \qw              & \octrl{1} & \qw \\
    & \qw          & \qw         & \qw      & \octrl{1} & \qw              & \ctrl{1}  & \qw \\
    & \targ{}      & \qw         & \qw      & \targ{}   & \qw              & \targ{}   & \qw
    \end{quantikz}
    \caption{Grouping 6-qubit gates into pairs to be combined, and the equivalent rewrite of the circuit that uses at most 5-qubit gates}
    \label{fig:circuit-optimization-task-solution}
\end{figure}

Figure~\ref{fig:circuit-optimization-task-solution} shows the solution steps for this task. First, one has to identify pairs of gates with similar control patterns. 
Then, each pair of gates with control patterns different in a single bit, for example, QA and TU pairs, can be replaced with one equivalent gate with fewer controls. 
In a more complicated scenario, the pair of gates that differ in two bits in a complementary pattern can be replaced by a gate with fewer controls surrounded with additional CNOT gates that capture this pattern (for example, the NM pair).

%=============================================================================
\section{Exploring quantum hardware}

Quantum hardware exploration assignments present both an opportunity and a challenge for the instructor. Very few quantum devices are currently available in the cloud, so the wait times for jobs to complete can be significant. Carefully planning the work on the assignment to get the job execution results back in time for the assignment deadline can be challenging for the students due to reasons unrelated to the quantum computing topics they are studying. 

In my course, I relied on local or cloud simulators for the majority of the assignments and used quantum hardware only for assignments that allowed students to gain insight into the noisy behavior of current quantum devices.

The basic assignments provided students with a simple pre-written quantum program and asked them to run it on a noiseless simulator and a real quantum device, comparing the results. The programs were as simple as a single-bit random number generator or a program that prepared a Bell state and measured both qubits. This approach introduced the students to the high-level concept of noise and its impact on program execution results even before a detailed discussion of noise in quantum systems.

In the quantum communication algorithms module, an assignment on quantum teleportation highlighted the limitations of the kinds of programs that can run on different devices. In Azure Quantum, these limitations are described as \textit{target profiles}. All cloud simulators and quantum devices available via Azure Quantum support either the ``No Control Flow'' profile, which allows the user to run only programs that return measurement results directly to the caller, or the ``Basic Measurement Feedback'' profile, which allows the programs to use the measurement results in simple conditional statements. The standard teleportation protocol cannot be executed on the targets with the ``No Control Flow'' profile, since they do not allow applying the fixup gates to the receiver state based on the sender's measurement results. Instead, the task explored a modification of the teleportation protocol that discarded the results of teleportation unless both sender's measurements yielded 0, that is, post-selected only the 25\% of the runs in which the receiver obtained the correct state without needing the fixup. The task itself did not focus on running the modified protocol on the quantum hardware devices; instead, it used the results from running it on a noiseless cloud simulator to prompt students to analyse the possible outcomes of the protocol mathematically and explain the simulation results.

The last assignment focused on exploring the size limitations of the programs that can be executed on a quantum device to produce a result better than a uniform random sample. The assignment provided the students with an implementation of Grover's search algorithm for finding alternating bit strings of the given length 0101... and 1010..., and asked them to run it on both the noiseless simulator and the quantum device for several problem sizes and explain the results.

\begin{figure*}
    \centering
    \includegraphics[width=.32\linewidth,valign=t]{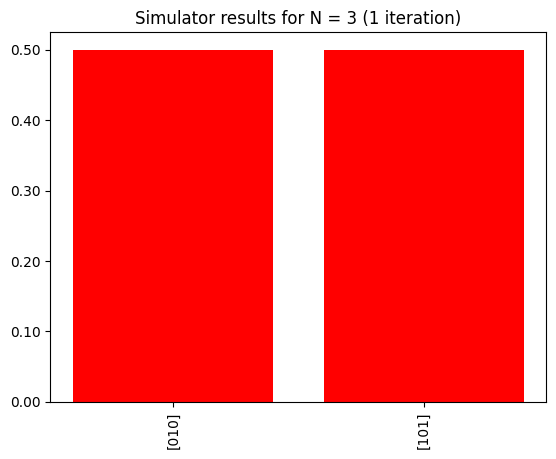}
    \includegraphics[width=.32\linewidth,valign=t]{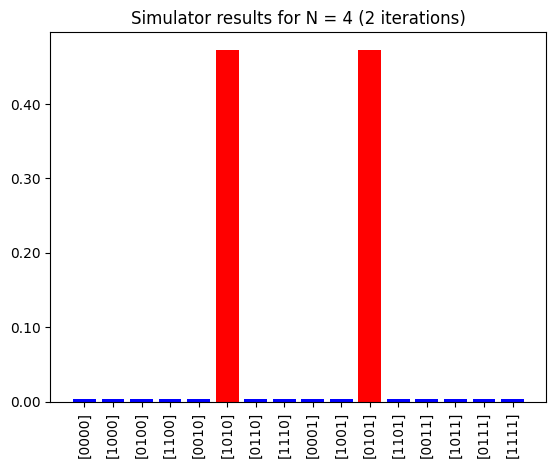}
    \includegraphics[width=.32\linewidth,valign=t]{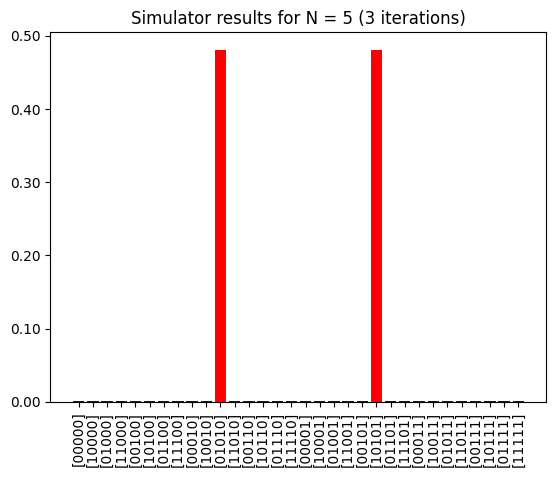}
    \\[\smallskipamount]
    \includegraphics[width=.32\linewidth,valign=t]{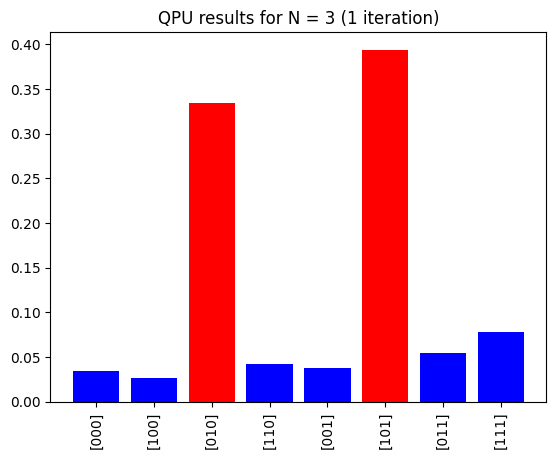}
    \includegraphics[width=.32\linewidth,valign=t]{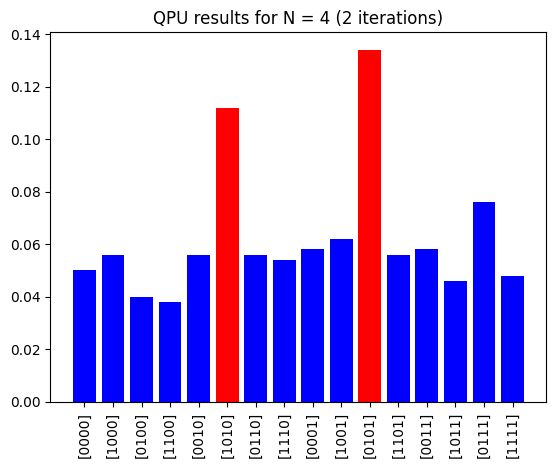}
    \includegraphics[width=.32\linewidth,valign=t]{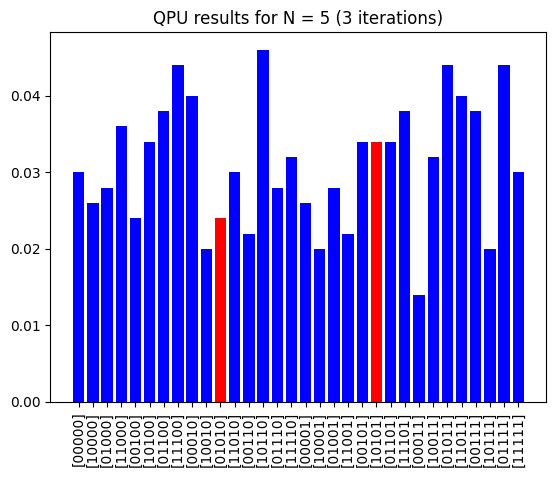}
    \caption{Example results of running the code for Grover's search algorithm on the simulator (top row) and on the quantum device (bottom row) for increasing bit string length $N$. Red bars mark the correct answers - states with alternating bits 0101... and 1010.... The results show that, while the success probability of the algorithm running on a simulator remains high as the bit string length increases, the success probability of the algorithm running on a quantum device decreases rapidly due to the increasing depth of the circuit executed and higher impact of noise.}
    \label{fig:grover-hardware-task}
\end{figure*}

Figure~\ref{fig:grover-hardware-task} shows an example of the results the students would get when running the code. They can observe that, in theory, the success probability of the algorithm remains high for all problem sizes. However, when the code runs on noisy hardware, solving larger problem instances requires deeper circuits, giving the noise more time to accumulate, to the point where it renders the algorithm results indistinguishable from noise.

%=============================================================================
\section{Final projects}

A final project (sometimes called a capstone project) is an independent exploration project offered at the end of the educational program that requires students to apply most of the skills they acquired during the program. 
This course used a final project instead of a final exam, as this aligned well with the course's focus on practical problem-solving exploration.

For the final projects, students worked individually to select and define a classical problem, develop an approach to solving it using Grover's search algorithm, implement the solution in Q\#, test and evaluate it, and present their work to the class.
The project included the following tasks:
\begin{enumerate}
    \item Implement an end-to-end solution to the problem and run it on a simulator to solve a small instance of a problem. The students could use one of the QDK simulators or the cloud simulators available via Azure Quantum.
    \item Solve multiple problem instances of different sizes on a simulator. This task shifts the focus from hardcoding the oracle for a single problem instance to using a more sophisticated implementation that coverts the problem definition into a matching oracle programmatically. 
    \item Test the solution, i.e., write the unit tests to check that the quantum oracle implemented in the solution is correct.
    \item Compare simulation speeds of the full state simulator and the sparse simulator available in the QDK when used to solve the same instances of the problem.
    \item Compare different solution approaches and the resources they require. For example, consider the problem of coloring the vertices of an $N$-vertex graph using three colors so that each pair of vertices connected by an edge is colored in different colors. One approach is to define the search space of all bit strings of length $2N$ that represent coloring $N$ vertices with colors 00, 01, 10, and 11, and use an oracle that checks both that the pairs of connected vertices have distinct colors and that none of the vertices are colored 11. Another approach limits the search space to only bit strings that represent coloring $N$ vertices with colors 00, 01, and 10 by changing the state preparation and reflection about the mean parts of the algorithm, and uses an oracle that only checks that the pairs of connected vertices have distinct colors. 
    
    Different approaches to problem solutions can have varying resource requirements. In the described example, the second approach reduces both the number of qubits required for the oracle implementation and the size of the search space compared to the first one, thus reducing the number of Grover iterations used. On the other hand, it increases the complexity of state preparation and reflection about the mean. The resources required for different implementations can be evaluated and compared using Azure Quantum Resource Estimator service.
    \item Run the solution for a small problem instance on one of the Azure Quantum hardware targets, and compare the results with the results produced by noiseless simulation.
    \item Use quantum counting to estimate the number of problem solutions. Outside of this task, the students could use a classical estimation of the number of solutions to get the optimal number of iterations to be used, or implement their solution without knowing the number of solutions.
    \item Explore automatic code generation for the same problem with Classiq platform\footnote{https://platform.classiq.io/}. The platform offered examples of code generation for Grover's search algorithm that could be modified to implement the students' projects.
    \item Present the results during the last lecture of the course. This encouraged students to practice their public speaking skills, get early feedback on their work, and learn more about the topics covered from other students' work.
\end{enumerate}

The complete solution to the problem and the final presentation were required for all final projects. The rest of the tasks were optional, and each students could select a subset of the tasks they wanted to focus on.

The breakdown of the project into specific tasks enabled standardized grading across the different approaches the students could've taken to solve and implement their problem of choice, since the variation of approaches within each task was significantly smaller than that within the whole projects.

%=============================================================================
\section{Lessons learned and future work}

Based on my observations teaching the course during the academic year 2022--23, the performance of 10 students who enrolled in one of the sessions during this period, and the student feedback collected after the course I believe that the described software-oriented approach is an effective way to introduce students with software development and engineering backgrounds to quantum computing. 

The students enjoyed the hands-on assignments and singled out the Quantum Katas as the part of the course that contributed the most to their learning.
Sharing the testing harnesses for the automatically graded programming assignments with the students led to higher success levels on these assignments compared to the earlier course session (academic year 2019--20), in which the students had to come up with ways to validate their solutions themselves.

On the other hand, several students struggled with the transition from the small guided assignments to the large project focused on self-driven problem solving and exploration. Of the 10 students enrolled, 2 failed the course, and one passed it on the second attempt. The students who received failing grades did not engage with the learning materials, including the weekly assignments, beyond the first weeks of the course.

Multiple students expressed special appreciation for the horizon-broadening nature of the course, and claimed that this course was among the best ones they have taken as part of their coursework.
One of the students even mentioned that they were interested in pursing quantum computing in their further graduate studies.

In the future offerings of this course, the homework assignments would benefit from the following improvements.

\begin{description}
\item [Offer more advanced debugging assignments.]
    Pure Q\# syntax errors turned out to be very easy to identify and fix. At the same time, weekly assignments did not offer students enough practice identifying and eliminating more sophisticated issues, such as the program containing elements that are not supported on certain hardware providers in Azure Quantum, that surfaced during their work on their final projects.

\item [Diversify hardware exploration assignments.]
    The most effective hardware exploration assignments were the ones that went beyond the mere observation of the existence of the noise, to exploring the impact of the noise on the feasibility of running end-to-end algorithms on the near-term quantum devices. It would be interesting, for example, to combine hardware exploration with circuit optimization, asking the students to optimize the given circuit to the point where the results of running it on quantum hardware are noticeably more accurate, or to explore whether running Grover's search for fewer iterations than the optimal iteration number would yield higher success probability due to smaller circuit depth.
    
\item [Offer multiple choices of the hardware provider.]
    Across the two sessions of the course, several assignments were negatively impacted by temporary unavailability of one of the providers' hardware targets. Enabling running each assignment on multiple hardware backends and allowing the students to choose the backend to use would mitigate the impact of such issues.
\end{description}
    
Finally, the artifacts created for these courses can be accessed by quantum computing educators worldwide to inspire them to adopt a similar teaching approach\footnote{https://github.com/microsoft/quantum-curriculum-preview}.

\section{Acknowledgements}  

I thank Kal Bugrara who invited me to teach the course ``CSYE6305: Introduction to Quantum Computing with Applications'' at Northeastern University, and the College of Engineering at Northeastern University for enabling this course. I am also grateful to my students for their valuable feedback.

\bibliographystyle{IEEEtran}
\bibliography{library,other}

\end{document}